\documentstyle[aps10,prb,preprint,aps,epsf,pstricks]{revtex}
\tightenlines 
\begin{document}
\setlength{\topmargin}{-0.5in}
\setlength{\oddsidemargin}{-0.3in}
\setlength{\textwidth}{6.5in}
\setlength{\textheight}{8.5in}
\newcommand{\bm}{\boldmath}
\title{\bf A molecular-dynamics  algorithm for mixed hard-core/continuous
potentials}
\author{{Yao A. Houndonougbo and Brian B. Laird} \\ {\small Department of Chemistry, University of Kansas, Lawrence, Kansas 66045}
 \\
{Benedict J. Leimkuhler} \\ {\small Department of Mathematics, University of Kansas, Lawrence, Kansas 66045}}
\date{\today}
\maketitle
\begin{abstract}
We present a new molecular-dynamics algorithm for integrating
the equations of motion for a system of particles interacting with
mixed continuous/impulsive forces. This method, which we call Impulsive Verlet, 
is constructed using operator splitting techniques similar to those that have
been used successfully to
generate a variety molecular-dynamics integrators. In numerical experiments,
the Impulsive Verlet method is shown to be  superior to previous methods
with respect to stability and energy conservation in long simulations. 
\end{abstract}
\newpage

\section{Introduction}

Purely collisional systems 
were among the first to
be studied by molecular-dynamics simulation\cite{Alder57}. 
These systems include hard spheres or
hard ellipsoids, which undergo purely elastic collisions, 
and square-well fluids, in
which an attractive impulse force at
a particular interparticle position is present in addition to the 
hard-core interactions. The algorithms
for such systems are exact, to within roundoff error, and consist of
free particle motion punctuated by exact resolution of the impulsive
collisions---the resulting phase space trajectory is discontinuous. 
On the other hand, the vast majority of current molecular-dynamics 
simulations are performed on systems with continuous potentials.
For such systems, the trajectory must be 
approximated using a numerical timestepping scheme
 such as the popular Verlet algorithm\cite{Verlet67}. 

There
exist, however, systems that are neither purely collisional,
nor continuous, but are  hybrids of the two. 
Important examples of such systems
are the restricted primitive model (RPM) for electrolyte solutions and 
dipolar hard spheres\cite{Hansen86}. In addition, the use of hard-core 
potentials
with attractive continuous tails is common in perturbative treatments
of liquids\cite{Weeks71}. Since the  the algorithms
for simulating impulsive and continuous systems are fundamentally
different from one another, the construction of hybrid methods for
mixed systems is non-trivial and little studied.   Consequently, the
vast majority of studies on such systems have utilized 
Monte Carlo simulation techniques, eliminating the possibility of
obtaining dynamical information.

In this paper, we present a new method for 
mixed hard-core/continuous potentials,
which we call the Impulsive Verlet algorithm.  
This algorithm is suitable for any continuous 
potential, is less likely than alternatives to miss collisions, 
and exhibits good stability and
energy conservation in long time simulation.  In the construction
of this new method we have been guided by recent work in the use
of Hamiltonian splitting methods for the development of efficient and
stable molecular-dynamics algorithms\cite{Tuckerman92,Sanz-Serna95}. 

A few  {\em ad hoc} hybrid methods have been  constructed 
for mixed (hard/soft) systems\cite{Stratt81,McNeil82,Heyes82,Suh90}.
All of these methods are rather similar, in that the 
particles are advanced according to the continuous forces by a time 
step using a standard algorithm for continuous potentials, usually some
variant of the Verlet algorithm, and the trajectories are checked for 
the existence of particle overlaps at the end or during the step. 
If no overlaps occur, the procedure is repeated for the next step. If 
overlaps (collisions) do occur, the system is returned to its state
before the step and then is advanced  without momentum
modification by the forces to the time of collision, and the
momenta are then modified according to the rules of elastic collision. This
process
is repeated until all collisions have been resolved
and the end of the time step is reached.  (One major difference between
the algorithms is whether overlaps are checked only at the end of
each step, or throughout the step. In the former case\cite{McNeil82}, it
is possible that glancing collisions are missed during the dynamics.) 
Heyes\cite{Heyes82} and Suh, {\it et al.}\cite{Suh90} apply such algorithms
to the restricted primitive model for electrolytes (hard-sphere with
embedded charges in a dielectric continuum) with some apparent
success.  Unfortunately, as with the other papers on algorithms
for mixed systems, no quantitative discussion on the stability or
accuracy of the algorithm is given, making it difficult to evaluate
the quality of the methods.

The Impulsive Verlet method is developed in the next two sections, 
followed by a discussion of certain
numerical experiments on two model systems,
comparing our scheme with the algorithm used in 
Suh,{\em et al.}\cite{Suh90}.

\section{Splitting Methods for Mixed Dynamics}

Consider a system of $N$ particles with instantaneous positions
$\mbox{\bm$q$} = (q_1,q_2,...,q_N)$ in $d$ dimensions 
interacting according to
a continuous potential $V_c(\{{\bf q_i}\})$, assumed for simplicity 
to be spherically symmetric and pairwise additive, that is,
\begin{equation}
V_{\rm c}({\bf q}) = \sum_{i=1}^{N} \sum_{j > i} \phi_{\rm c}(q_{ij}) \; ,
\end{equation}
where $q_{ij} \equiv \, \mid{\bf q_j} -  {\bf q_i}\mid$, and $\phi$ is
any smooth function of one variable.
In addition, suppose the particles to have a hard core of diameter 
$\sigma$; that is, when 
the distance between two particles is $\sigma$ an elastic collision
occurs that reflects the momentum of
each particle along the collision vector. Such a hard-sphere core 
can be represented formally by a discontinuous pair potential of the form
\begin{equation}
\phi_{\rm hs}(q_{ij})  = \left\{\begin{array}{cc}   \infty &  q_{ij}  \le \sigma, \\
               0, & q_{ij} > \sigma \;.
\end{array}
\right.
\label{hs_pot}
\end{equation}

We will define the energy function of the mixed system by analogy with
continuous dynamics as the sum of the kinetic and formal potential energies:
\begin{equation} H(\mbox{\bm$q$}, \mbox{\bm$p$}) = T(\mbox{\bm$p$}) + 
V_{\rm hs}(\mbox{\bm$q$}) + V_{\rm c}(\mbox{\bm$q$}), 
\label{hs_Hamil}
\end{equation}
where,
\[
V_{\rm hs} = \sum_{i=1}^N\sum_{j>i} \phi_{\rm hs}(q_{ij})
\]
and
\begin{equation}T(\mbox{\bm$p$}) = \frac{1}{2}\mbox{\bm$p$}^T \mbox{\bm$M$}^{-1}
 \mbox{\bm$p$},\label{kinetic} \end{equation}
is the kinetic energy ($\mbox{\bm$M$}$ is the mass matrix)
and $\mbox{\bm$p$}=(p_1,p_2,...,p_N)$, where each $p_i$
is a $d$-dimensional vector.  Despite appearances, 
this energy function is not, properly speaking,
a Hamiltonian.   Nonetheless, we can view the dynamics
of the hard-sphere fluid as the limiting dynamics in repulsive inverse-power
potentials 
of the form $V_{\rm sw}(r) = 1/r^\beta$, with $\beta$ a large positive integer.
In this sense and for the purpose of constructing numerical methods,
we can interpret the formal energy $H$ as representing
a very hard repulsive inverse-power  Hamiltonian.   We will often
refer to $H$ as 
the {\em pseudo-Hamiltonian}.

We define the flow map as the generator of the phase
space trajectory, 
\begin{equation}
\left(\begin{array}{c} {\bf q}(\tau+t) \\ {\bf p}(\tau+t) \end{array} \right)  =
\psi_{t,H} 
\left(\begin{array}{c} {\bf q}(\tau) \\ {\bf p}(\tau) \end{array} \right)  \; .
\end{equation}
The family of flow maps is closed under composition, 
\begin{equation} 
\psi_{t_1,H} \circ \psi_{t_2,H} =  \psi_{t_2,H} \circ \psi_{t_1,H} = \psi_{t_1 + t_2,H}, 
\label{concatenation}
\end{equation}
for any times $t_1$ and $t_2$. 

A continuous Hamiltonian system
can often be split into integrable 
subproblems with Hamiltonians $H_1$ and $H_2$\cite{Sanz-Serna95}:
\begin{equation} H(\mbox{\bm$q$}, \mbox{\bm$p$}) = H_1(\mbox{\bm$p$}) + H_2(\mbox{\bm$q$})\; .
\label{splitHamilt}
\end{equation}
The flow map of the full Hamiltonian can then be approximated as the 
concatenation of flow maps for the subproblems. There are a variety of 
ways of doing this, but the most common is based on a Trotter factorization
\begin{equation}
\psi_{h,H} = \psi_{\frac{h}{2},H_2} \circ \psi_{h,H_1} \circ \psi_{\frac{h}{2},H_2}
+ {\cal O}(h^3)\; , 
\label{trotter}
\end {equation}
where $h$ is the time step. 
For a separable Hamiltonian such as Eq.~\ref{hs_Hamil}
with $V_{\rm hs} = 0$, this factorization reduces to the usual velocity-Verlet
algorithm\cite{Swope82} when $H_1 = T({\bf p})$ and $H_2 = V_{\rm c}({\bf q})$.  

The splitting framework for continuous Hamiltonians
suggests a means of constructing integrators
for mixed impulsive/continuous systems.   A natural 
splitting for the pseudo-Hamiltonian is
to let $H_1 = T({\bf p}) + V_{\rm hs}({\bf q})$ and $H_2 = V_{\rm c}({\bf q})$. (Note that in
this case $H_1$ is a function of both ${\bf p}$ and ${\bf q}$, but since this 
represents a system with free particle motion punctuated by  elastic 
collisions, it is exactly integrable.)
This gives
\begin{equation}
\left(\begin{array}{c} {\bf q}^{n+1} \\ {\bf p}^{n+1} \end{array} \right)  =
\psi_{\frac{h}{2},V_c} \circ \psi_{h,T + V_{hs}} \circ \psi_{\frac{h}{2},V_c}
\left(\begin{array}{c} {\bf q}^n \\ {\bf p}^n \end{array} \right)  \; ,
\end{equation} 
where ${\bf q}^n$ and ${\bf p}^n$ are the approximations to the phase
space variables after the $n$-th  time step. 
In other words, the momenta are adjusted at the beginning of each time
step by one-half step according to the continuous forces (``kick''). 
The positions
are next advanced for one time step, resolving all elastic collisions, but
without further momentum modification by the continuous forces (``push'').
At the
end of the step, the momenta are advanced again by a half step using
the forces calculated from the new positions (another ``kick'').  
This is nearly identical to the
algorithm of Suh, {\em et al.}\cite{Suh90} except that there 
momenta are only defined at half steps and a
leap-frog formulation is used:
\begin{equation}
\left(\begin{array}{c} {\bf q}^{n+1} \\ {\bf p}^{n+1/2} \end{array} 
\right)_{\mbox{Suh}} =
 \psi_{h,T + V_{hs}} \circ \psi_{h,V_c}
\left(\begin{array}{c} {\bf q}^n \\ {\bf p}^{n-1/2} \end{array} \right)  \;.
\end{equation} 

Viewing the hard-sphere potential as being approximated by a very hard
inverse-power repulsive potential, we see that either of the above two 
splitting methods is
symmetric (i.e. time-reversible).  From Eq.~\ref{trotter}
we naively expect that such a
method (applied to the inverse-power potential approximation)
is second order accurate, meaning that in one step a {\em local error}
of size $O(h^3)$ is introduced; on a finite fixed time interval, these
errors accumulate, but the total growth
or {\em global error} is at most $O(h^2)$.   
However, 
the demonstration of third-order local error requires a $C^3$ solution, and
this assumption will break down in the limit of hard-sphere dynamics,
in particular during a collision step.  In fact, 
the local error introduced during 
a collision is really $O(h)$.   

We illustrate this point with the simple example of a nonlinear
``impact oscillator'' with one degree-of-freedom pseudo-Hamiltonian
\begin{equation}
H = \frac{p^2}{2} + \phi_{\rm hs}(q) + \phi_{\rm c}(q),
\label{h1deg}
\end{equation}
describing a point mass acted on by some
potential $\phi_c$ in collisional dynamics
with a hard wall at $q=\delta$.  The particle moves in the continuous
potential $\phi_{\rm c}$ according to Newton's equations, until an impact,
when $q=\delta$, then the momentum changes sign and the motion continues from
the impact point. 

Consider a numerical step
from the point $(q_0,p_0)$ at time $t=0$ 
for a timestep of size $h$ during which the
particle motion includes a single collision event.  (We mostly
use subscripts to index particle number and superscripts for timestep,
but for the discussion that follows we need to indicate powers of the momenta;
so for this one-particle model, we will use subscripts for the timestep index.)
We need to compute the local energy error contribution during a 
collisional step for this single degree-of-freedom model problem.
The sequence of computations is
\begin{eqnarray}
\hat{p} & = & p_0 - \frac{h}{2} \phi_{\rm c}'(q_0), \label{jump1}\\
h_{\#} & = & -\frac{q_0-\delta}{\hat{p}}, \label{jump2}\\
h_{\flat} & = & h-h_{\#}, \label{jump3}\\
\tilde{p} & = & - \hat{p}, \label{jump4}\\
q_1 & = & \delta - h_{\flat} \hat{p},  \label{jump5}\\
p_1 & = & \tilde{p} - \frac{h}{2} \phi_{\rm c}'(q_1)\; ,  \label{jump6}
\end{eqnarray}
where $h_{\#}$, and $h_{\flat}$ are the time to the next collision and
the time from that collision to the end of the time step, respectively.

Substituting the endpoint values into the energy relation, we quickly
find
\begin{eqnarray*}
\Delta H = H(q_1,p_1)-H(q_0,p_0) & = &\frac{1}{2}(\tilde{p} - \frac{h}{2} \phi_{\rm c}'(q_1)
)^2
+ \phi_{\rm c}(q_1) - \frac{1}{2} p_0^2 - \phi_{\rm c}(q_0),\\
& = & \frac{1}{2}(-p_0 + \frac{h}{2} \phi_{\rm c}'(q_0) - \frac{h}{2} 
\phi_{\rm c}'(q_1))^2\\
& & \hspace{0.3in}
+ \phi_{\rm c}(q_1) - \frac{1}{2} p_0^2 - \phi_{\rm c}(q_0).
\end{eqnarray*}
Expand $\phi_{\rm c}$ in a Taylor series about $q=\delta$, substitute, and 
cancel like terms to obtain
\begin{eqnarray*}
\Delta H  & = &   - \frac{h}{2} p_0 (\phi_{\rm c}'(q_0) -  \phi_{\rm c}'(q_1))
+ \frac{h^2}{8} (\phi_{\rm c}'(q_0) -  \phi_{\rm c}'(q_1))^2\\
& & \hspace{0.3in}
+ \phi_{\rm c}'(\delta)(q_1-q_0)  + \frac{1}{2} \phi_{\rm c}''(\delta)( (q_1-\delta)^2 - (q_
0-\delta)^2) + E_h.
\end{eqnarray*}
The remainder $E_h$ contains terms of order the third power of $h$ or
higher, i.e. $|E_h/h^3|$ is bounded for all $h<1$ such that the step contains
a collision.
Indeed, $\frac{h^2}{8} (\phi_{\rm c}'(q_0) -  \phi_{\rm c}'(q_1))^2$ is also of this
order, since $q_1-q_0$ is proportional to $h$.  From the equations
\[
q_1 = \delta - h_{\flat} \hat{p}, \hspace{0.5in}
q_0 = \delta - h_{\#} \hat{p},
\]
and the use of a Taylor series expansion of $\phi_{\rm c}'$, we arrive after discarding
terms of order three or higher at,
\[
\Delta H  =   \frac{h(h_{\#}-h_{\flat})}{2} \phi_{\rm c}''(\delta) \hat{p}^2
- \phi_{\rm c}'(\delta)(h_{\flat}-h_{\#})\hat{p}  + \frac{1}{2} \phi_{\rm c}''(\delta)
( h_{\flat}^2 - h_{\#}^2)\hat{p}^2 + \tilde{E}_h.
\]
with $\tilde{E}_h$ again of third order.
This finally leads to
\begin{eqnarray*}
\Delta H &  = &  \frac{(h_{\#}-h_{\flat})}{2} (h - (h_{\#}+h_{\flat})) \phi_{\rm c}''(\delta
) \hat{p}^2
- \phi_{\rm c}'(\delta)(h_{\flat}-h_{\#})\hat{p} + \tilde{E}_h\\
& = & - \phi_{\rm c}'(\delta)(h_{\flat}-h_{\#})\hat{p} + \tilde{E}_h.
\end{eqnarray*}

Therefore, the expected energy error introduced in this one collisional
step is 
\begin{equation} \label{ejump}
H(q_1,p_1)-H(q_0,p_0) = -\phi_{\rm c}'(\delta)(h_{\flat}-h_{\#})\hat{p} + \tilde{E}_h,
\end{equation}
where the quantity $|\tilde{E}_h/h^3|$ is
bounded independent of $h$.  (A similar
result would hold for the solution error.)

Technically speaking, it is incorrect to say that the energy jump in one step
is $O(h)$ since if we decrease the timestep $h$ sufficiently, there will
be no collision event within the particular step, and so the error will revert
to $O(h^3)$.
Nonetheless, in any timestepping simulation in which there are collision
events, these steps will introduce errors proportional to $h$.  If we define
the {\em local approximation error} $e_{loc}$
as the maximum of magnitudes of the local
errors introduced, then $e_{loc}$
is of first order in $h$,  
not third order as we would expect in the continuous case.
Since
there are, in general, a finite number of such collisions in any 
finite interval, the accumulation is bounded and 
the global error is also $O(h)$.    The apparent contradiction of an odd-order
symmetric method is just one of several anomalies that result from the 
complex transition from the smooth problem to the discontinuous limit.  
In another terminology, we could say that the splitting
method undergoes an {\em order reduction} for stiff potential wells.

From this discussion and Eq.~\ref{ejump}, we expect the
naive splitting method to give rather poor energy conservation, 
except in three special cases:
\begin{description}
\item[Case 1] Collisions do not occur {\em within}
timesteps but precisely {\em at} the timesteps, so third order is
recovered.
\item[Case 2] The collisions  occur at precisely the middle of
a timestep, so that the first order term in the error formula vanishes
and third order local energy drift is again recovered.
\item[Case 3] Third order will be recovered if
the {\em derivative of the continuous pair potential vanishes for 
two spheres in contact}.   
\end{description}
To illustrate this last point, we apply the method to one degree-of-freedom 
anharmonic ``impact oscillator" with a continuous potential, $\phi_c(q) = 
\frac{1}{2}q^2 + \frac{1}{4}q^4.$ We show in Fig.~\ref{orderdepWall} the 
maximum total energy error as a function of the time step when the wall 
is placed at $q = 0.00$ and $q = -4.00$.
One can see that the naive splitting is a second order method when the derivative 
at the wall vanishes . 

Because it is only applicable for a relatively limited class of
potentials the naive splitting method is not a candidate for a viable general
technique, however, it does provide a good starting point for the development
of a general method, which we call the Impulsive Verlet (IV) algorithm. 

\section{Impulsive Verlet}
To develop our method, we deliberately exploit two of the special cases in
the naive algorithm
for which third order can be expected, namely Cases 1 and 3 mentioned at the
end of the previous section. (Case 2, the situation that collisions
occur at the midpoint of the time interval, does not appear to be of
practical use.) We
begin by introducing an artificial splitting
of the continuous potential, $\phi_{\rm c}(q_{ij})$, into
into a short-ranged part, $\phi_1(q_{ij})$, and a 
long-range part, $\phi_2(q_{ij})$, according to
\begin{equation} 
\phi_{\rm c}(q_{ij}) = \phi_1(q_{ij}) + \phi_2(q_{ij})\;.
\label{splitPotential}
\end{equation}
(This decomposition is similar to that invoked in 
multiple timestepping\cite{Tuckerman92,Tildesley78,Windemuth91,Skeel94} 
molecular-dynamics algorithms.)
For the reasons discussed above, the long-range
(and therefore most expensive to 
calculate) part of the potential is defined so that the derivative vanishes 
at the hard-core separation.  We define $\phi_2(q_{ij})$ as follows:
\begin{equation} \phi_2(q) = \left\{ \begin{array}{ll} P(q_1), & q < q_1,
  \\
                              P(q), & q_1 \le q < q_2,  \\
                             \phi_c(q), & q \ge q_2\; ,
\label{V_2}
\end{array} \right.
\end{equation}
where, $q_1$ and  $q_2$ are parameters, and $ P(r) = A_o + A_1 r +A_2r^2 + A_3 r^3$ 
is a Hermite interpolant
introduced so that the two potentials are smooth to the order $C^1$ for any 
continuous potential. From  Eqns.~\ref{splitPotential}) and~\ref{V_2}), 
$\phi_1(q)$ is given by
\begin{equation} 
\phi_1(r_{ij}) =  \left\{ \begin{array}{cr} \phi_{c}(q) - P(q_1) & q < q_1 \\
                             \phi_{c}(q) - P(q) &q_1 \le q < q_2
,  \\
                              0 & q \ge q_2.
\end{array} \right.
\label{V_1}
\end{equation}
The continuity  condition,  $P(r_2) = \phi_{c}(r_2)$ , and  the smoothness 
conditions, $P^{\prime}(q_2) = \phi_c^{\prime}(q_2)$, $P^{\prime}(q_1) = 0$, 
and $P^{\prime \prime}(q_1) = 0$, allow us to calculate the coefficients 
of the Hermite interpolant, giving
\begin{equation} A_3 = \frac{\phi_c^{\prime}(q_2)}{6r_1(q_1-q_2)+3(q_2^2-q_1^2)}
 \; \mbox{for $q_1 \neq q_2$,} \label{A3}
\end{equation}
\begin{equation} A_2 = -3 q_1 A_3, \label{A2} \end{equation}
\begin{equation} A_1 = 3 q_1^2 A_3, \label{A1} \end{equation}
\begin{equation} A_0 = -(A_1 q_2 + A_2 q_2^2 + A_3q_2^3) + V_{c}(q_2)\;.
\label{A0} 
\end{equation}
(An example of this potential splitting for an inverse-sixth-power attractive 
potential, $\phi_c(q)=-\epsilon(\sigma/q)^6$,  is shown in 
Fig.~\ref{splitPotInv6_1.1_1.2}.)

Next, we define $N$-body potentials $V_1$ and $V_2$ as a sum of pair contributions from $\phi_1$ and $\phi_2$, respectively.
We then split the total Hamiltonian in the following way:
\begin{equation} H_1(\mbox{\bm$q$} ,\mbox{\bm$p$}) =  T(\mbox{\bm$p$}) + 
V_{\rm hs}(\mbox{\bm$q$}) + V_1(\mbox{\bm$q$}) \label{H1exp} 
\end{equation}
and
\begin{equation} H_2(\mbox{\bm$q$})  = V_2(\mbox{\bm$q$})\;. 
\label{H2exp} 
\end{equation}

The Trotter factorization (Eq.~\ref{trotter}) is now applied to this
splitting. The problem now is that $H_2$ is not integrable and its flow
map must be approximated. This is done is the following way:
\begin{equation}
\psi_{H_2,h} \approx \prod_{i=1}^{n_c+1} \psi_{V_{\rm hs}}\circ \psi_{V_1,\tau_i^{(c)}/2} 
\circ \psi_{T,\tau_i^{(c)}} \circ \psi_{V_1,\tau_i^{(c)}/2} \;,
\end{equation}
where $n_c$ is the number of hard-sphere collisions between during the time
step $h$, $\tau_i^{(c)}$ is the time between each collision (with $\tau_1^{(c)}$ being
measured from the beginning of the time step until the first collision 
and $\tau_{n+1}^{(c)}$ measured from the last collision to the end of the time 
step), and $\psi_{V_{\rm hs}}$ is an 
operator representing the resolution of each elastic collisions. This is 
essentially the 
execution of a Verlet step of length $\tau^{(c)}$ between each elastic collision.
The collision times can be calculated since the Verlet step generates a quadratic
trajectory, which together with the collision condition  for two particles $i$ and $j$
can be written as
\begin{equation} 
\|\mbox{\bm$q$}_i(\tau^{(c)}) - \mbox{\bm$q$}_j(\tau^{(c)})\|^2 - 
\sigma^2 = 0,\label{coll_cond} \;
\end{equation}
generates a quartic equation for $\tau^{(c)}$. 

We describe below the algorithm for the Impulsive Verlet molecular-dynamics 
simulation in more detail.

\vspace{0.2in}
\begin{center}
\fbox{
\begin{minipage}[t]{4.5in}

\footnotesize

\begin{center}
\underline{Impulsive Verlet Timestepping Algorithm}
\end{center}
\noindent
\begin{itemize}
\item[] ${\bf p}_i^{n+1/2,0} = {\bf p}^{n,0} + \frac{1}{2} 
           {\bf F}_{2,i}({\bf q}^{n,0}) h$
\item[] do $i_c = 1,n_c$ 
\begin{itemize}
\item[] ${\bf p}_i^{n+1/2,i-1/2} = {\bf p}^{n+1/2,i-1} + \frac{1}{2} 
           {\bf F}_{1,i}({\bf q}^{n,i-1}) \tau_c^{i}$
\item[] ${\bf q}^{n,i} = {\bf q}^{n,i-1} + {\bf M}^{-1} 
{\bf p}^{n+1/2,i_c} \tau_c^{i}$
\item[]$\tilde{\bf p}_i^{n+1/2,i} = {\bf p}^{n+1/2,i-1/2} + 
\frac{1}{2} {\bf F}_{1,i}({\bf q}^{n,i}) \tau_c^{i}$
\item[]${\bf p}^{n+1/2,i} = \psi_{V_hs} \left( \begin{array}{c}
{\bf q}^{n,i_c}\\ \tilde{\bf p}^{n+1/2,i} \end{array} \right )$
\end{itemize}
\item[] end do
\item[]${\bf p}_i^{n+1/2,n_c+1/2} = {\bf p}^{n+1/2,n_c} + \frac{1}{2} 
        {\bf F}_{1,i}({\bf q}^{n,n_c}) (h - \sum_{i=1}^{n_c} \tau_c^{i})$
\item[] ${\bf q}^{n+1,0} = {\bf q}^{n,n_c} + {\bf M}^{-1} 
{\bf p}^{n+1/2,n_c+1/2} (h - \sum_{i=1}^{n_c} \tau_c^{i})$ 
\item[]${\bf p}_i^{n+1/2,n_c+1} = {\bf p}^{n+1/2,n_c+1/2} + \frac{1}{2} 
{\bf F}_{1,i}({\bf q}^{n+1,0}) (h - \sum_{i=1}^{n_c} \tau_c^{i})$ 
\item[]${\bf p}_i^{n+1,0} = {\bf p}^{n+1/2,n_c} + \frac{1}{2} 
           {\bf F}_{2,i}({\bf q}^{n+1,0}) h$
\end{itemize} 

\end{minipage}
}
\end{center}
\vspace{0.2in} 

To make sure that no collisions are missed it is necessary to ensure
that the quartic equation (Eq.~\ref{coll_cond}) is accurately solved to give the  
nearest root to zero. This is not a trivial problem as the solution
becomes increasingly unstable as smaller time steps  are used (i.e. when the
time to collision is small). To ensure the inaccuracies are not large
enough to affect the overall accuracy and order of the method, we employ
Laguerre's method\cite{NumRec} to find all roots of the quartic and take
the smallest, positive real root, which is then refined using Newton-Raphson.
This proved to be sufficient at all but the very smallest time steps studied.

There is a small probability that the Impulsive Verlet method can miss a
grazing collision, since the trajectories that are followed in
determining collisions are quadratic approximations.  However, this probability
is greatly reduced in comparison to the method of Suh, {\it et al.} or any other algorithm
that uses linear motion to determine the collisions.

\section{Numerical Experiments}
We test the Impulsive Verlet algorithm using as our continuous potentials,
$\phi_c(q)$, the Lennard-Jones  potential
\begin{equation}
\phi_{c,LJ} = 4\epsilon \left [\left (\frac{\sigma}{q} \right )^{12} - 
\left (\frac{\sigma}{q} \right )^6 \right ] \; . 
\end{equation}
and an attractive inverse-sixth-power potential
\begin{equation}
\phi_{c,6} = -\epsilon \left (\frac{\sigma}{q} \right )^6 \; . 
\end{equation}
In both potentials $\sigma$ is the same as the hard-core diameter.
We truncate both potentials at the distance $q^*_c = q_c/\sigma = 2.5$ and, 
to ensure their continuity, they are shifted so that the value of the
potential at the cutoff is zero. 
In implementing the Impulsive Verlet algorithm, 
we split each potential as prescribed in  Eq.~\ref{splitPotential}-~\ref{A0},
with $q_1$ and $q_2$ as input parameters. 
For the Lennard-Jones potential there is, of course, a natural splitting,
namely that of Weeks, Chander and Anderson (WCA)\cite{Weeks71}, where the
potential is split at the minimum with $q_1^* = q_2^* = 2^{1/6}$, which gives the
following splitting: 
\begin{equation} \phi_{1,LJ}(q;\mbox{WCA}) =  \left\{ \begin{array}{cc} 4\epsilon
 [ (\frac{\sigma}
{q})^{-12} - (\frac{\sigma}{q})^{-6}]
+ \epsilon  & q < 2^{\frac{1}{6}}\sigma, \\
                              0, &q \ge 2^{\frac{1}{6}}\sigma.
\end{array} \right.
\label{V_1LJ}
\end{equation}
\begin{equation} \phi_{2,\mbox{LJ}}(q,\mbox{WCA}) = \left\{ \begin{array}{cc} -\epsilon, & 
q< 2^{\frac{1}{6}}\sigma ,  \\ 
4\epsilon [ (\frac{\sigma}{q})^{-12} - (\frac{\sigma}{q})^{-6}] , & q \ge 2^{\frac{1}{6}}\sigma,
\end{array} \right.
\label{V_2LJ}
\end{equation}

The MD simulations were carried out on systems of 108 particles.
The system of reduced units was 
chosen so that all quantities are dimensionless. So, as units of distance and 
energy we used the potential parameters $\sigma$ and $\epsilon$, respectively, 
and the mass of one atom as the unit mass. The unit of time is 
$(m \sigma^2/\epsilon)^{1/2}$. An asterisk superscript indicates reduced units.
Except were otherwise indicated all simulations are performed using a reduced density 
$\rho^{\ast} = \rho \sigma^3 = 0.9$ and reduced temperature $T^{\ast} = kT/\epsilon = 2.5$.
In addition,  a cubic box with periodic boundary conditions is used.
For greater efficiency, the MD program incorporates three neighbor lists \cite{Allen87} for the evaluation
of the short-range force, the long-range force, and the collision times. 

The results of the Impulsive Verlet on the instantaneous total energy for the 
Lennard-Jones and the attractive inverse sixth continuous potentials are 
illustrated in Fig.~\ref{LJ_ImpVlet_Suh_4E-3} and \ref{Inv6_ImpVlet_Suh_4E-3}.
A  comparison to the naive splitting algorithm of 
Suh, {\it et al.}\cite{Suh90} is also made for both potentials.
The superiority in energy conservation and stability of
the Impulsive Verlet algorithm over the naive splitting method is
striking.  

We  study in Fig.~\ref{ljOrder} the order of the method 
while varying $q_1$ and $q_2$. The order is
obtained by plotting (on a log-log) the maximum energy error for a 
fixed-length simulation versus the time step. A comparison with a straight 
line of slope two  tells us that the method is of second order for 
various values of $q_1^*$ and $q_2^*$. (Note the slight deviation of the slope at
very small time steps from the theoretical value of 2.0 is due to the 
difficulty in solving the quartic equation for the collision times when
the time to collision is very small. This is not a real problem in practice 
since the goal of molecular-dynamics simulation is to use the largest 
time steps possible.)

Finally, to demonstrate the ability of the Impulsive-Verlet  method 
to yield relevant dynamical quantities, we show in Fig.~\ref{vAuto} the result for the normalized velocity 
autocorrelation function , 
$ C(t) = \langle {\bf v}(t)\cdot{\bf v}(0)\rangle / \langle {\bf v}(0)\cdot{\bf v}(0)\rangle$,  
for the Lennard-Jones system (108 particles)  with $\rho^* = 0.9$ and $T^* = 0.9$. In this calculation
we use a splitting with $q_1^* = 1.122$ and $q_2^* = 1.5$. 

\section{Conclusion}
We have introduced a molecular-dynamics method for mixed hard-core/continuous 
potentials, which we refer to as the Impulsive Verlet algorithm.
This algorithm is produced by extending general potential splitting methods 
to the specific case of mixed potentials.  In addition to providing a 
mechanism for generating the Impulsive Verlet method, the potential 
splitting formalism helps to understand the failings of previous methods.
The Impulsive Verlet algorithm uses a 
quadratic trajectory between collisions and does not miss any collisions of
the approximate trajectory.
As a result the algorithm is suitable for any type of continuous potential, is 
second order, has good energy preservation, and is far more 
stable over long time simulation than previously integrators for such systems.
(A detailed theoretical analysis of the algorithm is the subject of current research.)

\section{Acknowledgements}
The authors were supported in this work by NSF Grant DMS-9627330. In addition,
the simulations reported herein were performed on computers provided by
the Kansas Institute for Theory and Computational
Science (KITCS) and the Kansas Center for Advanced Scientific Computing
(KCASC).
The authors thank Steve Bond for helpful discussions.

\bibliography{master}
\bibliographystyle{laird_notitle}

\newpage
\begin{figure}
\caption{\normalsize{The maximum total energy error as a function of the time step for
one degree-of-freedom anharmonic ``impact oscillator" is  using the naive splitting approach.
The wall is placed at $q = -4.00$ (square) and at $q = 0.00$ (circle).
Comparison is made with lines of slope two(solid line) and one(dashed line).
}}
\label{orderdepWall}
\end{figure}
\begin{figure}
\caption{\normalsize{ A potential splitting of a inverse-sixth-power attractive interaction,
$-(\frac{\sigma}{q})^6$, with $q_1/ \sigma = 1.1$ and $q_2/\sigma = 1.200$. The
short range and the long potentials are (a) and (b) respectively.}}
\label{splitPotInv6_1.1_1.2}
\end{figure}

\begin{figure}[h]
\caption{\normalsize{Instantaneous total energy for a 108 particle simulation using a Lennard-Jones
continuous potential with a hard-sphere core, using both  Impulsive Verlet
(solid line) and the naive splitting algorithm of Suh, {\it et al.}
(dashed line). The time step is $h^* = 4 \times 10^{-3}$.} }
\label{LJ_ImpVlet_Suh_4E-3}
\end{figure}
\begin{figure}[h]
\caption{\normalsize{Instantaneous total energy for a 108 particle system interacting
via an inverse sixth-power attractive potential with a hard-sphere core,
using both Impulsive Verlet(solid line) and  the naive splitting algorithm
of Suh, {\it et al.} (dashed line). The time step is $h^* = 4\times 10^{-3}$ .} }
\label{Inv6_ImpVlet_Suh_4E-3}
\end{figure}
\begin{figure}[h]
\caption{\normalsize{The maximum total energy error as a function
of the time step for a system
of 108 particles using the Impulsive Verlet algorithm and the Lennard-Jones
potential.  Comparison is made with a line of slope two.}}

\label{ljOrder}
\end{figure}
\begin{figure}[h]
\caption{\normalsize{Normalized velocity autocorrelation as a function of time for
108 Lennard-Jones particles at $\rho^* = 0.9$ and $T^* = 0.9$, using
the Impulsive Verlet algorithm with a time step $h^* = 1\times 10^{-2}$.
}} 

\label{vAuto}
\end{figure}

\newpage


\begin{figure}
\epsfysize = 12 cm
\epsfxsize =\textwidth
\centerline{
\epsfbox{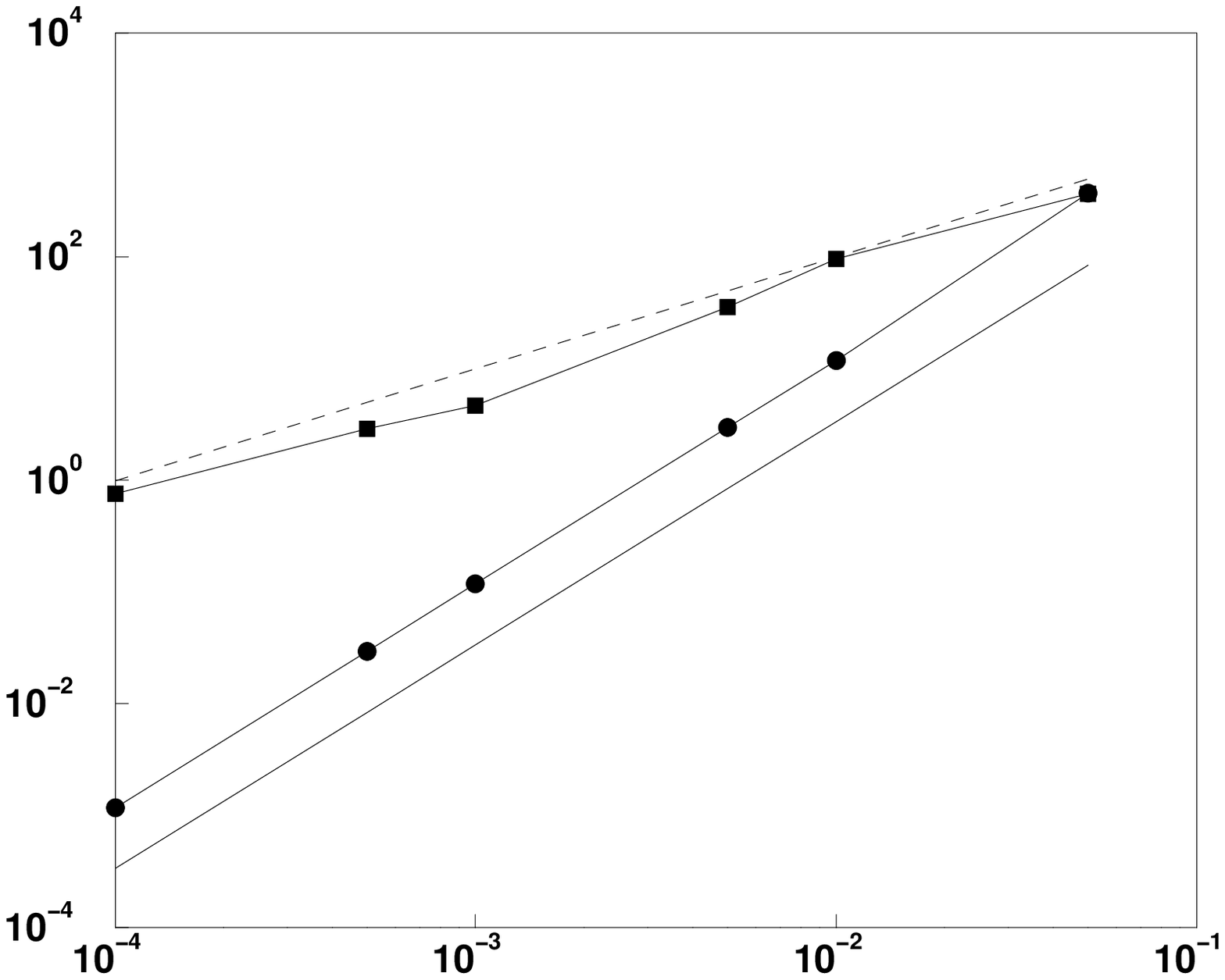}
}
\rput[l]{90}(0.0,6.8) {\Large{$\delta E$}} 
\rput[l]{0}(9.0,0.3) {\Large{$h$}}
\end{figure}
\newpage
\begin{figure}
\epsfysize = 10 cm
\epsfxsize =\textwidth
\centerline{
\epsfbox{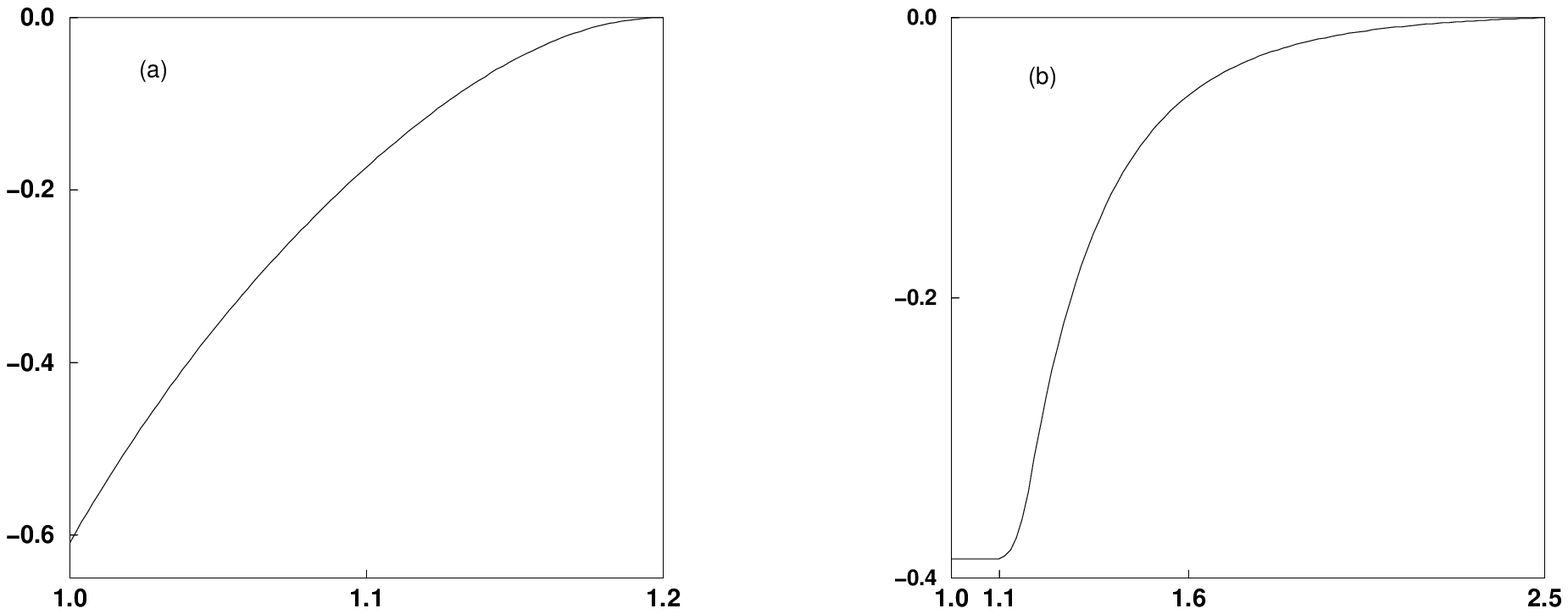}
}
\rput[l]{90}(0.0,4.8) {\Large{$V_{\rm c}(r_{ij})/ \epsilon$}}
\rput[l]{90}(9.2,4.8) {\Large{$V_{\rm c}(r_{ij})/ \epsilon$}}
\rput[l]{0}(13.2,0.2) {\Large{$r/\sigma$}}
\rput[l]{0}(4.2,0.2) {\Large{$r/\sigma$}}
\end{figure}
\newpage
\begin{figure}[h]
\epsfysize = 12 cm
\epsfxsize = \textwidth
\epsfbox{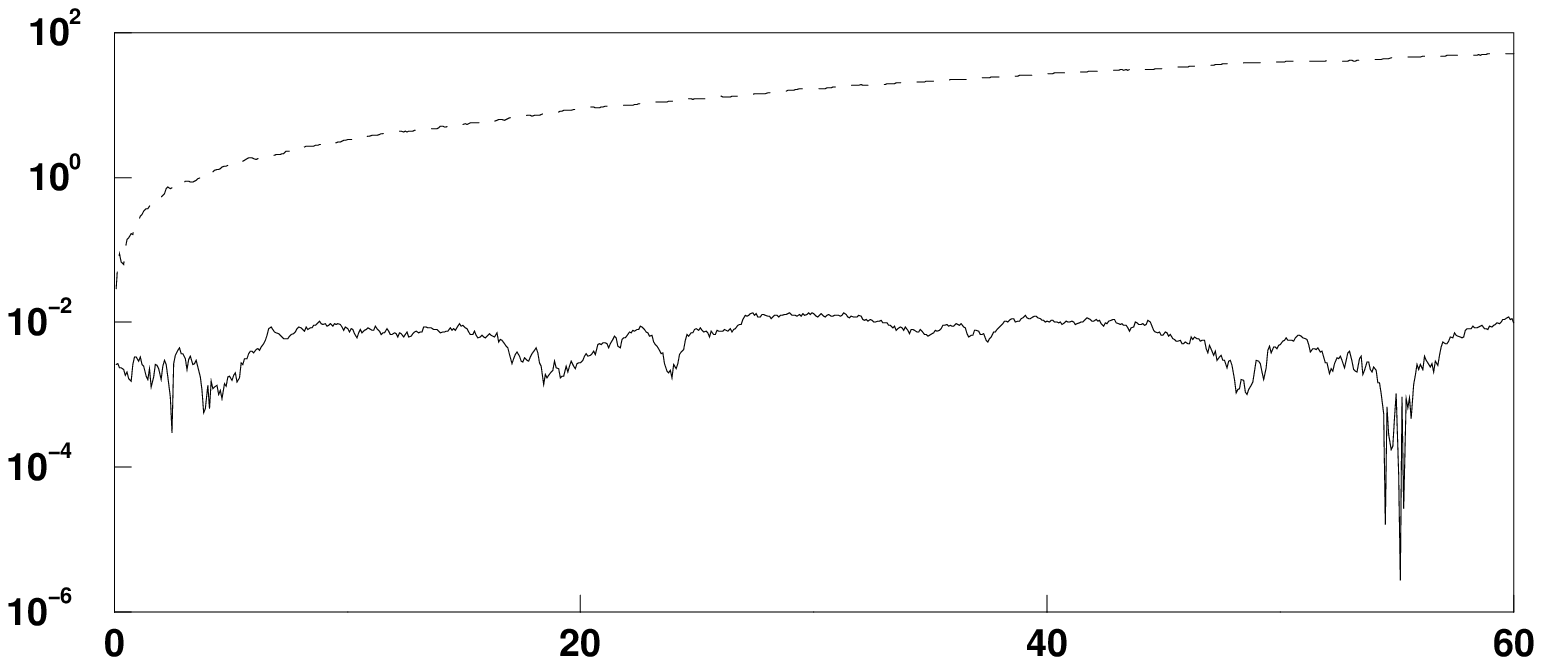}
\uput[r]{0}(-1.2,8.2) {\Large{$\delta E^*$}} 
\uput[r]{0}(8.0,0.3) { \Large{$t^*$}}
\end{figure}
\newpage
\begin{figure}[h]
\epsfysize = 12 cm
\epsfxsize = \textwidth
\epsfbox{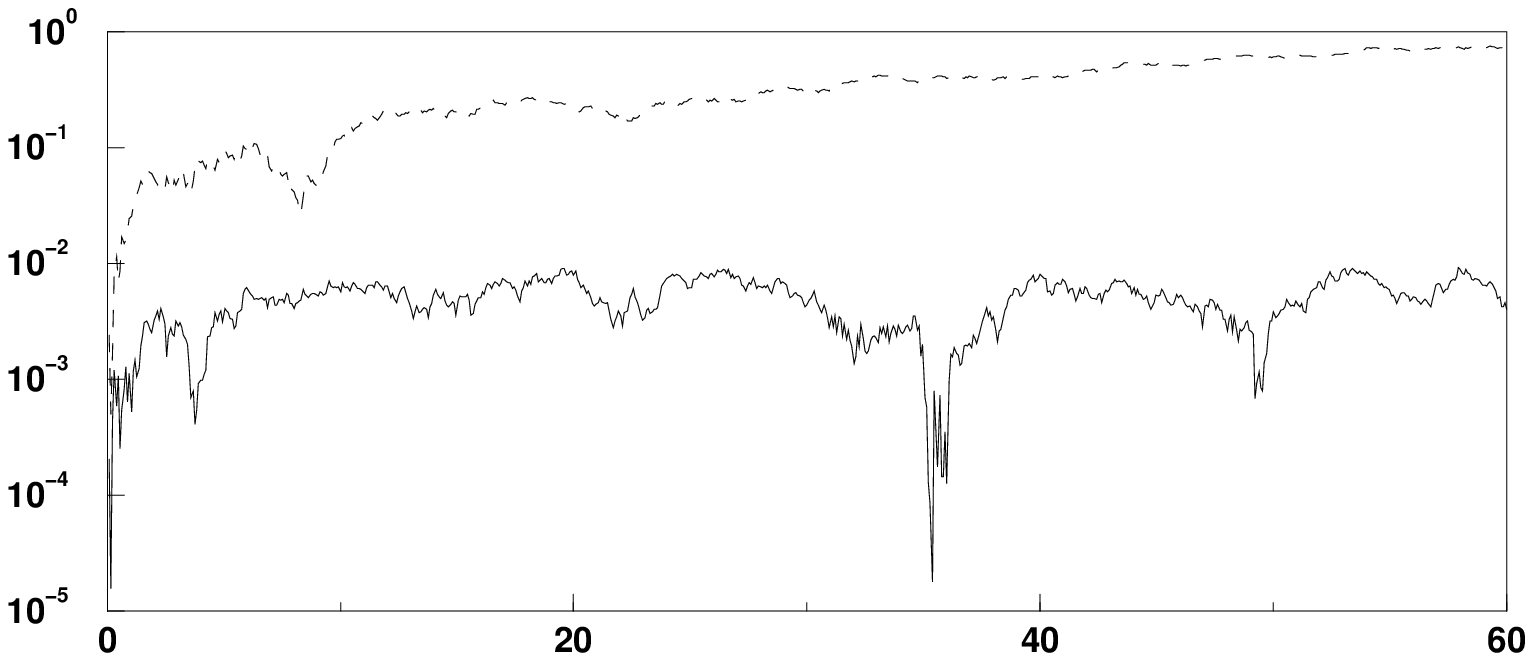}
\uput[r]{0}(-1.2,7.0) {\Large{$\delta E^*$}} 
\uput[r]{0}(8.0,0.3) { \Large{$t^*$}}
\end{figure}
\newpage
\begin{figure}[h]
\epsfysize = 12 cm
\epsfxsize = \textwidth
\centerline{
\epsfbox{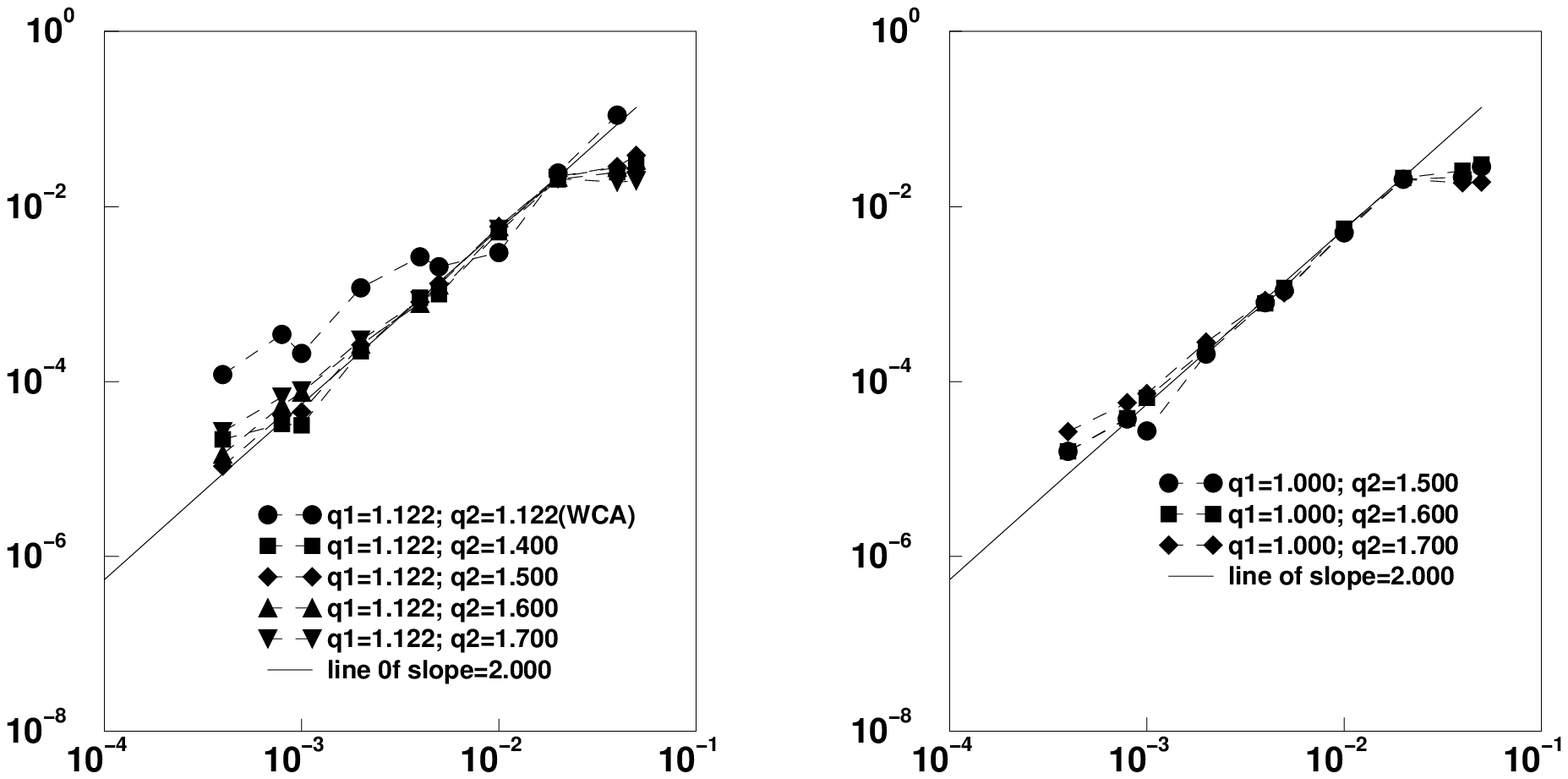}
}
\rput[l]{90}(0.0,6.5) {\Large{$\delta E^*$} }
\rput[l]{90}(8.8,6.5) {\Large{$\delta E^*$} }
\rput[l]{0}(4.8,0.2) {\Large{$h^*$} }
\rput[l]{0}(12.8,0.2) {\Large{$h^*$} }
\end{figure}
\newpage
\noindent
\begin{figure}[h]
\epsfysize = 12 cm
\epsfxsize = \textwidth
\centerline{
\epsfbox{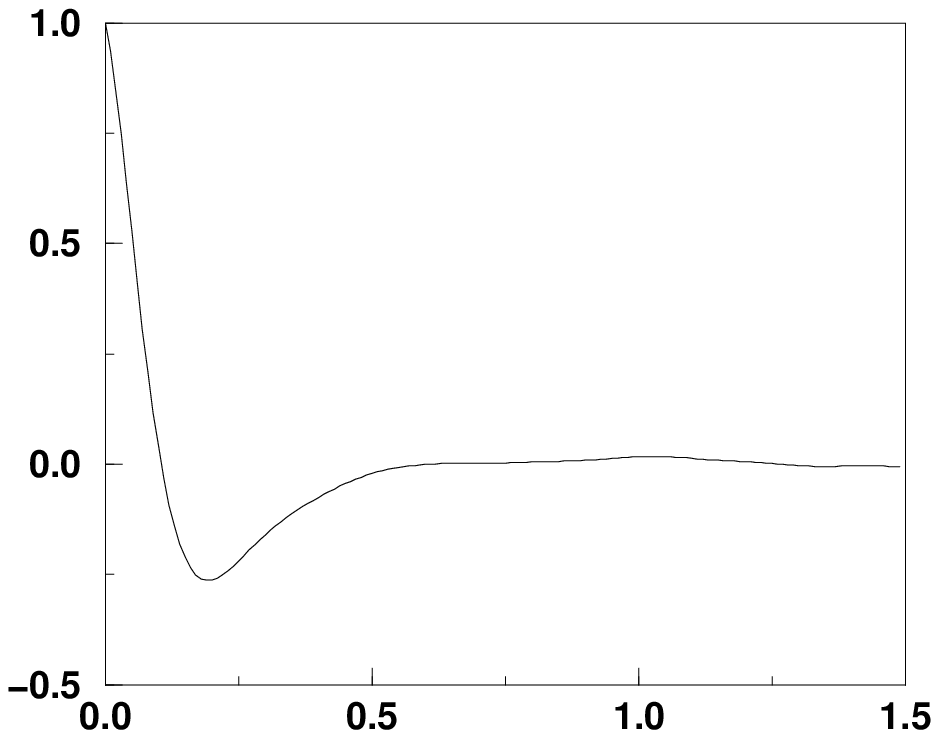}
}
\rput[l]{90}(0.0,6.8) {\Large{C(t)}}
\uput[r]{0}(9.0,0.2) { \Large{$t^*$}}
\end{figure}

\end{document}